\documentclass[12pt]{article}

\usepackage{scicite}
\usepackage{times}
\usepackage{siunitx}
\usepackage{graphicx}
\usepackage{float}
\usepackage{array}
\usepackage{caption}
\usepackage{booktabs}

\newenvironment{sciabstract}{%
\begin{quote} \bf}
{\end{quote}}

\title{Perforated red blood cells enable compressible and injectable hydrogels as therapeutic vehicles} 

\author
{Oncay Yasa$^{1}$, Fikru M. Tiruneh$^{1}$, Miriam Filippi$^{1}$,\\
Aiste Balciunaite$^{1}$, Robert K. Katzschmann$^{1\ast}$\\
\\
\normalsize{$^{1}$Soft Robotics Laboratory, Department of Mechanical and Process Engineering,}
\\
\normalsize{ETH Zurich, Tannenstrasse 3, 8092 Zurich, Switzerland}\\
\\
\normalsize{$^\ast$To whom correspondence should be addressed; E-mail: rkk@ethz.ch.}
}

\date{}

\begin{document}
\baselineskip24pt 
\maketitle 

\begin{sciabstract} 
    Hydrogels engineered for medical use within the human body need to be delivered in a minimally invasive fashion without altering their biochemical and mechanical properties to maximize their therapeutic outcomes. In this regard, key strategies applied for creating such medical hydrogels include formulating precursor solutions that can be crosslinked \textit{in situ} with physical or chemical cues following their delivery or forming macroporous hydrogels at sub-zero temperatures via cryogelation prior to their delivery. Here, we present a new class of injectable composite materials with shape recovery ability. The shape recovery is derived from the physical properties of red blood cells (RBCs) that are first modified via hypotonic swelling and then integrated into the hydrogel scaffolds before polymerization. The RBCs’ hypotonic swelling induces the formation of nanometer-sized pores on their cell membranes, which enable fast liquid release under compression. The resulting biocomposite hydrogel scaffolds display high deformability and shape-recovery ability. The scaffolds can repeatedly compress up to $\sim$\num{87}\% of their original volumes during injection and subsequent retraction through syringe needles of different sizes; this cycle of injection and retraction can be repeated up to ten times without causing any substantial mechanical damage to the scaffolds. Our biocomposite material system and fabrication approach for injectable materials will be foundational for the minimally invasive delivery of drug-loaded scaffolds, tissue-engineered constructs, and personalized medical platforms that could be administered to the human body with conventional needle-syringe systems.
\end{sciabstract}


\subsection*{Introduction}
Injectable materials can serve as functional medical platforms to perform minimally invasive operations inside the human body~\cite{ifkovits2010injectable, shibata2010injectable, purcell2014injectable, chao2020smart, wang2021injectable}. Most of the injectable materials used for this purpose rely on \textit{in situ} physical or chemical crosslinking of precursor solutions following their administration~\cite{Yu2008, Overstreet2012, Mathew2018}. However, this strategy can cause health issues due to undesired leakage and off-target distribution of precursor solutions within the body prior to their polymerization. Therefore, a safer approach involves fabricating fully-cured injectable medical platforms outside the body, and subsequently delivering them to the desired sites of action in a minimally invasive manner with conventional medical systems, \textit{e.g.}, needle-syringe. A pre-formed injectable medical platform should: (i) be compressed and delivered under pressure; (ii) maintain its biochemical and mechanical integrity during administration; and (iii) recover its original shape after the administration to perform its designated medical activity.

Cryogelation is a conventional approach to transform pre-formed hydrogels into injectable medical platforms~\cite{bencherif2012injectable}. Cryogels are macroporous hydrogel scaffolds that can be created by crosslinking certain materials below the freezing point of water (\textit{i.e.}, sub-zero temperatures)~\cite{gun2013cryogels, henderson2013cryogels}. Because of their macroporous nature that allows rapid liquid exchange, cryogels can tolerate reversible deformation at high strain and display shape memory ability, namely a capability of geometric and volumetric recovery following their injection through conventional needle-syringe systems~\cite{eggermont2020injectable}. Due to these peculiar properties, cryogels have emerged as a viable option for minimally invasive delivery of therapeutic scaffolds to areas in the body that are difficult to reach noninvasively via conventional medical approaches~\cite{razavi2019three, memic2019latest, ccimen2021injectable}. So far, cryogels have mostly been used for cancer immunotherapy, cell delivery, and tissue engineering applications~\cite{savina2007cryogels, koshy2014injectable, shih2018injectable, yuan2021injectable, najibi2022cryogel}. However, the stochastic pore formation in cryogels and the limited number of materials that can be polymerized at sub-zero temperatures hamper their widespread adoption as pre-formed injectable medical platforms. Consequently, it is valuable to explore alternative strategies that can be used for inducing the formation of macropores within polymer networks, which can enable rapid liquid exchange with the surrounding environment, and thus applied for creating injectable hydrogels in a simpler and faster manner.

Red blood cells (RBCs) are anucleated cells with no organelles. As they circulate within the body, their stability is ensured by the properties of their cell membrane and their distinctive oval biconcave disk shape. RBCs can deform and pass through tiny capillaries (the smallest having a few micrometer diameters) during circulation~\cite{meram2013shear}. The deformability of the RBCs results from: (i) the structural properties of the horizontal cytoskeletal components; (ii) the vertical interactions between the cytoskeleton and integral transmembrane complexes; and (iii) the resistance of the cytoplasmic pool that derives from the hydration state of the cells, intracellular viscosity, and surface-volume interaction~\cite{huisjes2018squeezing}. The biomechanical properties, abundance, biocompatibility, biodegradability, and non-immunogenicity of RBCs make them an ideal candidate to create medical platforms for therapeutic delivery applications inside the human body~\cite{villa2016red, glassman2021red, li2021clinical, wang2022erythrocyte}. In recent years, these properties of the cells have also led to the development of advanced therapeutic delivery platforms with multifunctional operation capabilities for active targeting and on-demand cargo release~\cite{wu2014turning, wu2015rbc, alapan2018soft, buss2020nanoerythrosome}. However, these medical platforms mostly rely on the biophysical performance of individual cell and there is still no therapeutic platform that benefits from the mechanical properties of the RBCs as a bulk within functional hydrogel scaffolds for minimally invasive concentrated drug delivery.

Here, we present a strategy that utilizes the physical properties of the RBCs to fabricate injectable hydrogel scaffolds as medical platforms. While realizing our injectable platform, we most importantly modify the RBCs with a hypotonic solution that creates nanometer-sized pores on the cell membrane via damage due to hypotonic swelling. The nano-perforated membrane elicits a rapid exchange of liquids between the internal cell compartment and the external environment, which is crucial to consistent volumetric cell deformation in conditions of mechanical stress. When embedded in a hydrogel matrix, the perforated RBCs act as space-holders within the hydrogel network and undergo fast volumetric variations, causing the resulting scaffold to deform under compression. Our investigations with different conventional needle-syringe systems show that these hydrogel scaffolds can compress up to $\sim$\num{87}\% of their original volumes while injection and can be potentially used for concentrated therapeutic delivery to the intended sites of action in a minimally invasive manner. Moreover, the stability of the scaffolds allows us to retract them without rupture by using different conventional needle-syringe systems following their operation. Finally, the robustness of these scaffolds enables them to be re-injected up to ten times without any significant damage. Overall, we are confident that this injectable material engineering strategy will lead to the creation of personalized centimeter-scale hydrogel scaffolds that can be used for minimally invasive drug delivery, tissue engineering, and regenerative medicine applications inside the human body.

\clearpage
\subsection*{Results \& Discussion}
\paragraph*{Formulation of injectable hydrogel scaffolds.}
Injectable hydrogel scaffolds are composed of perforated RBCs that are incorporated into their polymer networks during crosslinking reactions. RBCs isolated from murine whole blood were initially modified with a sequential hypotonic-isotonic treatment procedure \textbf{(Fig.~1A)}. By having a lower osmotic pressure than the cytoplasmic content of the RBCs, the hypotonic solution initiates cell swelling due to water intake into the cytoplasm~\cite{stewart2018intracellular}. While swelling, nanometer-sized pores start to form on cell membranes. These nanopores allow the cytoplasmic content of the cells (\textit{i.e.}, mainly intracellular hemoglobin) to be released to the liquid medium via passive diffusion. The average size of the nanopores and the recovery of cells after undergoing this treatment process are determined by the osmolality of the hypotonic solution and the duration of cell incubation within it. More specifically, if the hypotonic treatment process is mild enough, the RBCs can reseal the formed nanopores on their cell membranes during the isotonic treatment process and recover their original biconcave disk shapes. However, if the osmolality of the hypotonic solution is below a certain threshold (\textit{i.e.}, \num{2}:\num{1} mixed dH$_2$O:PBS solution in this case) and the cell incubation period is longer than \num{10} min at room temperature, the cells cannot fully recover themselves during the isotonic treatment process and form permanent spherical architectures with unsealed nanopores on them. In such a case, the isotonic treatment process is applied only to prevent the full burst of the cells for obtaining the desired stable perforated RBCs.

Once the perforated RBCs were obtained, a photo-reactive prepolymer solution was formulated to fabricate injectable hydrogel scaffolds. The prepolymer solution was consisted of \num{3} wt\% gelatin methacryloyl (GelMA), \num{1} wt\% photoinitiator (\textit{i.e.}, LAP), and $\sim$\num{.5d9} perforated RBCs/mL. While preparing the prepolymer solution, the pH of the GelMA-LAP mixture was adjusted to 7.4 before mixing the cells to prevent their burst and degradation at acidic pH conditions. In the next step, different square-shaped hydrogel scaffolds were obtained by pouring the prepolymer cell-mixed solution into 3D-printed molds and crosslinking them with ultraviolet light (\textit{i.e.}, \num{405} nm) for a minute. Finally, the injectability of the fabricated hydrogel scaffolds through different conventional needle-syringe systems was examined for finding the maximum compressibility of the scaffolds for their minimally invasive administration \textbf{(Fig.~1B)}.

\paragraph*{Mechanism of injectability.}
The perforated RBCs act as “space holders” within the hydrogel matrix as the crosslinking reaction stabilizes the polymer network around them. Under compression, liquids inside these cells escape into the local environment through the nanopores created on their cell membranes during the hypotonic treatment. The cell shrinkage allows sufficient macroscopic deformation for the hydrogel to compress through a small dimension needle. Once injected, the collapsed RBCs reabsorb liquid from the surrounding environment and swell to their initial sizes, causing the hydrogel to regain its original shape and display shape memory behavior \textbf{(Fig.~1C)}.

\paragraph*{Characterization of syringe injectability.}
Initially, the injectability of \qtyproduct{10 x 10 x 0.5} {\mm} pre-formed hydrogel scaffolds through \num{14} G needles, which have \num{1.75} \si{\mm} inner diameters, was examined to determine the perforated RBC concentration necessary for the administration of the scaffolds through conventional needle-syringe systems without rupture \textbf{(Fig.~2A)}. This investigation revealed that $\sim$\num{.5d9} perforated RBCs per milliliter of prepolymer solution provide enough compressibility to the scaffolds for injection through needle-syringe systems. Below this concentration, the polymer network starts to dominate the scaffolds' composition, and thus, the scaffolds do not have enough perforated RBCs for sufficient liquid exchange with their surroundings under compression. In the opposite case, the perforated RBCs start to hinder the crosslinking of GelMA and eventually prevent hydrogel scaffold formation.

The injectability of pre-formed hydrogel scaffolds through needles having different inner diameters was tested to find out the smallest needle size that can be used for their minimally invasive administration to the human body. Hydrogel scaffolds with four different sizes (\textit{i.e.}, \qtyproduct{10 x 10 x 0.5} {\mm}, \qtyproduct{7 x 7 x 0.5} {\mm}, \qtyproduct{5 x 5 x 0.5} {\mm}, and \qtyproduct{4 x 4 x 0.5} {\mm}) were fabricated. Afterward, these scaffolds were injected through \num{14}, \num{16}, \num{18}, and \num{20} G needles; having inner diameters of \num{1.75}, \num{1.35}, \num{0.97}, and \num{0.64} \si{\mm}, respectively \textbf{(Fig.~2B)}. Syringes with a centric cone were used for the injection to homogeneously distribute the forces acting on the scaffolds while they were entering into the needles. The injection speed was manually controlled to avoid inducing high shear stress while the scaffolds were passing through the needles. A chamber filled with buffer was chosen as the target destination while testing the scaffolds’ injectability to avoid collision with solid structures or surfaces that could cause mechanical disruption by impact. Failure to meet these conditions caused the scaffolds to rupture during the injection. Lastly, the syringes and the needles were not coated with any material that would prevent interactions between their surfaces and the scaffolds. This examination revealed that the largest scaffold dimensions for safe injection through \num{14}, \num{16}, \num{18}, and \num{20} G needles were \qtyproduct{10 x 10 x 0.5} {\mm}, \qtyproduct{10 x 10 x 0.5} {\mm}, \qtyproduct{10 x 10 x 0.5} {\mm}, and \qtyproduct{5 x 5 x 0.5} {\mm}, respectively \textbf{(Fig.~2C \& Video S\ref{vid:video_s1})}.

For validating the importance of perforated RBCs inside hydrogels for injectability, \qtyproduct{10 x 10 x 0.5} {\mm} hydrogel scaffolds containing either unmodified RBCs or composed of only \num{3} wt\% GelMA were injected through \num{14} G needles (\textit{i.e.}, the needles having the largest inner diameter). These tests demonstrated that the scaffolds had substantial mechanical damage and did not show the ability to recover their initial morphology following injection \textbf{(Fig.~2D, Video S\ref{vid:video_s2} \& S\ref{vid:video_s3})}. Hence, we show the necessity of having perforated RBCs inside hydrogel networks for achieving high compressibility and successful injectability without rupture.

\paragraph{Mechanical characterization of injectable hydrogel scaffolds and estimation of their compression percentage.}
The injectable hydrogel scaffolds and the control group that did not contain any RBCs were analyzed with a tensile testing machine to determine how the perforated RBCs alter the mechanical properties of the scaffolds. The maximum tensile strain of the injectable hydrogel scaffolds was drastically smaller compared to the control group (\textit{i.e.}, $\sim$\num{110}\% strain for the injectable hydrogels and $\sim$\num{180}\% strain for the control group) \textbf{(Fig.~2E)}. This material behavior may be due to a disruption of the amount of crosslinked groups in the hydrogel network when the perforated RBCs are incorporated. The analyses of Young's moduli showed that the injectable hydrogel scaffolds had a slightly smaller but comparable Young's modulus to the control group (\textit{i.e.}, \num{5.5} and \num{6.4} kPa, respectively), which supports a difference in the degree of crosslinking of the two groups.

The compression percentage of the injectable hydrogel scaffolds was calculated by assuming they occupy the full radial space of the needles during injection and that they elongate up to 110\% of their original length without any significant volume change. We assumed that the hydrogels formed a cylinder with the radius of the needle gauge and elongated length based on a side length of the hydrogel \textbf{Fig.~2F}. For example, a \qtyproduct{10 x 10 x 0.5} {\mm} hydrogel scaffold (\textit{i.e.}, the largest scaffold examined in this work), which has a volume of \num{50} mm$\textsuperscript{3}$, can pass through a volume of \num{24} to \num{38} mm$\textsuperscript{3}$ according to its elongation state while injecting with a \num{14} G needle. Therefore, this scaffold should compress between \num{25} to \num{52}\% of its original volume for injection from \num{14} G needles without any mechanical damage and rupture. The calculations also revealed that this largest scaffold should compress between \num{77} to \num{85}\% of its original volume for passing through \num{18} G needle. These results show that the scaffolds can not be properly injected through any needles if they have to be compressed more than \num{87}\% of their original volume \textbf{(Table 1)} These compression percentages are similar to the values achieved by the other injectable pre-formed hydrogel platforms.

\begin{table}[H] 
    \centering
    \begin{tabular}{ c c c c c}
         \toprule
         \textbf{Gauge} & \textbf{Scaffold 1} & \textbf{Scaffold 2} & \textbf{Scaffold 3} & \textbf{Scaffold 4} \\
         \midrule
         \textbf{14 G} & n.c. & 0 - 4\% & 0 - 31\% & 25 - 52\% \\ 
         \textbf{16 G} & 0 - 30\% & 12 - 43\% & 37 - 60\% & 56 - 72\% \\
         \textbf{18 G} & 44 - 64\% & 55 -71\% & 68 - 79\% & 77 - 85\% \\
         \textbf{20 G} & 75 - 84\% & 80 - 87\% & n.i. & n.i. \\
         \bottomrule
    \end{tabular}
    \caption{Compression percentages of injectable hydrogel scaffolds while injection with different needles (\textit{i.e.}, \num{14}, \num{16}, \num{18}, and \num{20} G). The dimension of the Scaffold 1, 2, 3, and 4 are \qtyproduct{4 x 4 x 0.5} {\mm} (8 mm$^3$), \qtyproduct{5 x 5 x 0.5} {\mm} (12.5 mm$^3$), \qtyproduct{7 x 7 x 0.5} {\mm} (24.5 mm$^3$), and \qtyproduct{10 x 10 x 0.5} {\mm} (50 mm$^3$), respectively. n.c.; no compression \& n.i.; no injection.}
    \label{tab:table_1}
\end{table}

\paragraph*{Retraction and reusability of injectable hydrogel scaffolds.}
Once administered into a liquid medium, the hydrogel scaffolds could be also retracted with conventional needle-syringe systems without introducing any macroscopic damage \textbf{(Fig.~3A, Video S\ref{vid:video_s4})}. For instance, \qtyproduct{10 x 10 x 0.5} {\mm} scaffolds could be easily retracted with \num{14} and \num{16} G needles. The retraction of the scaffolds was an effortless process, with no precautions to be taken, apart from placing the syringes close enough for proper suction. To the best of our knowledge, no retraction of such kind for hydrogel scaffolds has been ever documented. By shrinking, the perforated RBCs allow the surrounding polymer structure to deform in order to fit through the needle. Inside the syringe chamber, the RBCs re-uptake the liquid from their surroundings and swell to their original size, leading the scaffolds to regain their original shapes. Retractability could be promising for those medical applications in which scaffolds have to be withdrawn after their use within the body.

In addition to the retractability, the scaffolds were also robust and durable; for instance, the largest one could be injected up to ten times through \num{14} G needles without any noticeable damage or fracture under microscopic inspection but it could be injected only up to five times through a \num{16} G needle. Scaffolds were imaged after each injection test to show their structural integrity \textbf{(Fig.~3B)}. The same construct could not be re-injected from a \num{18} G needle \textbf{(Fig.~3C)}. Therefore, there could be a possible shear dependent effect that limits the reusability of injectable hydrogel scaffolds. In this context, the size variations of the perforated RBCs inside hydrogel networks was assessed via scanning electron microscopy, revealing that cell size decreased after injection possibly due to cell membrane damage during compression and their partial recovery following injection \textbf{(Fig.~3D)}. These size variations of the perforated RBCs after injection process could be the primary reason behind the decrease in their number of re-injection according to the used needle size.

\paragraph*{Drug release under compression.}
Following injectability and reusability investigations, concentrated therapeutic delivery with this injectable platform under compression was examined by administering \qtyproduct{5 x 5 x 0.5} {\mm} scaffolds through \num{14}, \num{16}, \num{18}, and \num{20} G needles. The \qtyproduct{5 x 5 x 0.5} {\mm} scaffolds were used for these experiments because they are the largest scaffolds that can be injected through all four needles with a certain compression percentage. A model drug molecule (\textit{i.e.}, FITC-Dextran) was physically entrapped into the polymer network during the fabrication process and its release to the liquid environment from hydrogel scaffolds after injection was examined with fluorescence spectroscopy. The results showed that the scaffolds release their therapeutic content according to the amount of compression \textbf{(Fig.~3E)}. However, the drug release from these scaffolds under compression can potentially be reduced by the covalent modification of polymer networks with therapeutics.

\paragraph*{Injectability and retractability of complex geometries.} 
The injectability and retractability of different geometries were tested to further present the mechanical stability of the injectable hydrogel scaffolds \textbf{(Video S\ref{vid:video_s5})}. First, it was demonstrated that cubic hydrogel scaffolds can be easily administered by using conventional needle-syringe systems when the perforated RBCs are incorporated into their polymer networks. In this regard, \qtyproduct{3 x 3 x 3} {\mm} hydrogel scaffolds were injected and retracted with a \num{14} G needle \textbf{(Fig.~3F)}. While passing through the needle, the scaffolds were compressed up to $\sim$\num{73}\%. Following the injection process, they immediately recovered their original volumes without any detectable macroscopic damage. Next, a large hydrogel scaffold of \num{1} mm thickness and with edges prone to damage was successfully retracted and injected without any detectable mechanical damage by using a \num{14} G needle \textbf{(Fig.~3G)}. This result demonstrated that larger volumes of pre-formed hydrogels can be easily administered to the desired sites of actions with conventional needle-syringe systems when it will be necessary.

\subsection*{Conclusion \& Future Perspectives}
Minimally invasive concentrated therapeutic delivery could aid us in fighting diseases at hard-to-reach body locations. Prior to this work, cryogels were the only viable option for administering pre-formed centimeter-sized hydrogel scaffolds through conventional needle-syringe systems for minimally invasive concentrated therapeutic delivery.~\cite{ccimen2021injectable, eggermont2020injectable} Here, we presented an alternative strategy to create syringe-injectable pre-formed hydrogel scaffolds by integrating perforated RBCs into hydrogel networks. This injectable material formulation opens perspectives in controllable materials for future biomedical applications.

In the future, this bio-integrated material will serve to create injectable platforms for minimally invasive and personalized treatment procedures by using patient-specific, safe, and soft materials (such as patient-derived RBCs and biocompatible polymers). Such a platform could potentially reduce the immune response in medical applications.~\cite{ceylan2019translational} The herein presented hydrogel scaffolds could be retracted with needle-syringe systems, which opens the possibility of using deformable materials as temporary biomedical systems. For instance, medical platforms steered by external magnetic fields could use these retractable scaffolds to potentially alleviate the biocompatibility issues of the required toxic magnetic particles.

Future investigations will focus on tuning the mechanical properties of the fabricated hydrogel scaffolds by varying the sizes of the incorporated RBCs. In fact, the RBC diameters in mammalians range from $\sim$\num{4} to \num{8} µm depending on their origins. By using different RBC sizes in specific areas of the hydrogel network, the mechanical properties of scaffolds can be tuned locally. These different formulations could be combined to obtain sophisticated 3D constructs with complex deformation behaviors. For instance, different formulations of matrices with incorporated perforated RBCs could serve as bio-inks that can be selectively 3D printed to form heterogeneous functional constructs.

\subsection*{Materials \& Methods}
\paragraph*{Synthesis of gelatin methacryloyl.}
The primary amine groups of gelatin from porcine skin (gel strength \num{300} g Bloom, Type A) were transformed into methacryloyl groups (a mixture of methacrylamide and methacrylate groups) through its reaction with methacrylic anhydride~\cite{loessner2016functionalization}. Briefly, we first dissolved \num{10} wt\% gelatin in phosphate-buffered saline (PBS) at \num{50} \si{\degreeCelsius}. Then, we added \num{2.4}\% (v/v) methacrylic anhydride into this solution under continuous stirring and let it to react with gelatin for \num{1} h at \num{50} \si{\degreeCelsius}. After that, we dialyzed the reaction mixture against ultrapure water for \num{3} d with \num{12}-\num{14} kDa cutoff dialysis tubing. Finally, we lyophilized the samples for \num{3} d and stored them in dark at \num{4} \si{\degreeCelsius} until further usage.

\paragraph*{Preparation of perforated red blood cells.}
Red blood cells (RBCs) were isolated from murine whole blood (Swiss Webster, Taconic Biosciences, Germany) with a density gradient. The density gradient was created inside a \num{15} mL centrifuge tube by using three different Percoll solutions that were \num{21}\%, \num{38}\%, and \num{80}\% (v/v) Percoll in PBS. Initially, \num{5} mL of \num{80}\% Percoll solution was placed into the centrifuge tube. Then, \num{3} mL of \num{38}\% and \num{2} mL of \num{21}\% Percoll solutions were carefully transferred on top of the first one in a sequential manner. After that, \num{1} mL of whole blood, which was one-to-one diluted with PBS, was gently put on top without mixing the solutions. Finally, the sample was centrifuged at \num{4} \si{\degreeCelsius} and \num{800} \textit{g} for \num{20} min and dark red middle phase containing the RBCs were transferred into a new vial for further usage.

\paragraph*{} After isolating the RBCs, they were modified as previously described with a slight modification~\cite{alapan2018soft, buss2020nanoerythrosome}. Briefly, the cells were washed three times with PBS, and then they were incubated in a hypotonic solution (\textit{i.e.}, \num{1}:\num{2} mixed PBS:dH$_2$O solution) at \num{4} \si{\degreeCelsius} by shaking for \num{10} min. Afterward, the sample was centrifuged at \num{4} \si{\degreeCelsius} and \num{2000} \textit{g} for \num{5} min, and then the pellet containing the modified RBCs was re-suspended in an isotonic solution (\textit{i.e.}, PBS). After that, the sample was incubated at \num{37} \si{\degreeCelsius} for \num{15} min inside this isotonic solution. Finally, the sample was centrifuged at room temperature (RT) and \num{2000} \textit{g} for \num{5} min, and the pellet (\textit{i.e.}, perforated RBCs) was dissolved in PBS and stored at \num{4} \si{\degreeCelsius} until further usage.

\paragraph*{Formulation of composite material.}
The photoreactive polymer solution was composed of \num{3} wt\% gelatin methacryloyl (GelMA), \num{1} wt\% lithium phenyl-2,4,6-trimethylbenzoylphosphinate (LAP), and $\sim$\num{.5d9} perforated RBCs suspended in \num{1} mL of PBS. Initially, GelMA and LAP were completely dissolved in PBS by vortexing and sonication at \num{60} \si{\degreeCelsius} for \num{30} min in a dark condition. Then, the pH of this GelMA-LAP solution is adjusted to \num{7.4}. Next, the solution was heated up to \num{37} \si{\degreeCelsius} in a water bath for \num{30} min. In parallel, the perforated RBCs were diluted with PBS according to the desired final concentration (\textit{i.e.}, $\sim$\num{.5d9} cells/mL), and then they were centrifuged at RT and \num{2000} \textit{g} for \num{5} min. Finally, the pellet containing the perforated RBCs was homogeneously mixed in the GelMA-LAP solution by vortexing, and the prepolymer solution was stored at \num{4} \si{\degreeCelsius} and in a dark condition until further usage.

\paragraph{Fabrication of injectable hydrogel scaffolds.}
Injectable hydrogel scaffolds were fabricated by molding the photoreactive polymer solution into various 3D-printed structures (negative molds), and then polymerizing it with a \num{405} nm ultraviolet (UV) light source for \num{1} min. Shortly, the negative molds were fabricated with a stereolithography (SLA) 3D printer (Form 3+, Formlabs, Somerville, MA) from Clear resin. After fabrication, the unreacted resin residues were washed in an isopropanol bath for \num{10} min, and then the molds were fully cured at \num{60} \si{\degreeCelsius} for \num{30} min by applying \num{405} nm UV light. After having the 3D-printed structures, the prepolymer solution was carefully molded into them while avoiding air bubbles, and then it was crosslinked by applying a \num{405} nm UV light for \num{1} min. Finally, the crosslinked samples were removed from their molds inside PBS with a scalpel, and then their injectability was tested with conventional needle-syringe systems.

\paragraph*{Characterization of injectable hydrogel scaffolds.}
Conventional three-part needle-syringe systems were used to examine the injectability of the fabricated hydrogel scaffolds. Syringes with a volume of \num{10} or \num{20} mL were used with needles having different gauge sizes that were \num{14}, \num{16}, \num{18}, and \num{20} G. The syringes had a centric cone with a Luer-Lock connector, and both the syringes and the needles did not have any inner chemical coatings. Before experiments, the syringes were initially filled with approximately \num{5} mL of PBS, and then the hydrogel scaffolds were carefully loaded into the syringes with a spatula. After submerging into the solution, the hydrogel scaffolds detached from the spatula and sank towards the centric cone of the syringes. Finally, the scaffolds were injected into a beaker filled with PBS, and the videos were recorded with a Sony APS-C ILCE 6000 digital camera having a Sony E PZ 16-50 mm F3.5-5.6 OSS objective (Sony Corporation, Tokyo, Japan). For retraction experiments, the hydrogel scaffolds were first transferred to a beaker filled with PBS. After that, the needle-syringe systems were filled with $\sim$\num{1} mL of PBS to reduce the friction. Finally, the scaffolds were sucked into the syringes through the needles by applying negative pressure. For re-usability experiments, the hydrogel scaffolds were not retracted through the needles by applying negative pressure, but they were rather transferred into the needle-syringe systems each time with the help of a spatula.

\paragraph*{Mechanical characterization of hydrogels.}
The elastic modulus \si{(\unit{\exa})} and strain at the break of the injectable hydrogel scaffolds were examined with a uniaxial tensile testing machine (Instron 5942, Instron, Norwood, MA). Rectangular negative molds (\qtyproduct{11 x 4 x 0.5} {\mm}) were fabricated with an SLA 3D printer from the Clear resin. After that, samples composed of the photoreactive polymer solution (\textit{i.e.}, \num{3} wt\% GelMA and \num{1} wt\% LAP) with or without the perforated RBCs were molded into the 3D-printed structures, and then, crosslinked with a \num{405} nm UV light for \num{1} min. Next, the samples were removed from their molds inside PBS and they were immediately mounted to the testing platform to avoid drying. Finally, the samples were characterized at a strain rate of \num{0.05} per min up to rupture and break, and their elastic moduli were calculated from the slope of the elastic deformation region.

\paragraph*{Scanning electron microscopy.}
The injectable hydrogel scaffolds were fixed with \num{2.5}\% (v/v) glutaraldehyde in PBS solution at \num{4} \si{\degreeCelsius} for \num{15} min and then rinsed three times with PBS. After that, they were dehydrated in a series of ethanol solutions having increasing concentrations of alcohol (\textit{i.e.}, \num{20}\%, \num{40}\%, \num{60}\%, \num{80}\%, and \num{100}\% (v/v)) for \num{3} min in each solution. Next, dehydration was continued through a series of hexamethyldisilazane (HMDS) - ethanol solutions, with increasing HMDS concentrations (\textit{i.e.}, \num{33}\%, \num{66}\%, and \num{100}\% (v/v)) for \num{30} min in each solution. Then, the samples were air-dried overnight in a chemical fume hood and they were coated with \num{10} nm of gold using a spin coater (Leica EM ACE 600, Leica Microsystems, Wetzlar, Germany). Finally, they were examined with a Zeiss Merlin scanning electron microscope (Carl Zeiss Inc., Oberkochen, Germany) using an accelerating voltage of \num{3} keV and an in-lens detector.

\paragraph*{Drug encapsulation and release.}
Drug release from \qtyproduct{5 x 5 x 0.5} {\mm} hydrogel scaffolds upon their injection through \num{14}, \num{16}, \num{18}, and \num{20} G needles was examined with a UV-Vis spectrophotometer (Infinite 200 PRO, Tecan, Switzerland). FITC-Dextran (fluorescein isothiocyanate dextran) dissolved in PBS (\num{1} mg/mL, \num{20} kDa) was used as a model drug molecule and encapsulated into the injectable hydrogel scaffolds by directly adding into the polymer solution (\textit{i.e.}, \num{3} wt\% GelMA with \num{1} wt\% LAP containing $\sim$\num{.5d9} perforated RBCs/mL) before its photo-crosslinking with a \num{365} nm UV light. Initially, the absorbance of \num{1} mg/mL FITC-Dextran at \num{490} nm was investigated; \num{490} nm wavelength is the optimal absorbance for this model drug molecule. After that, \num{1} mg/mL FITC-Dextran was exposed to \num{365} nm UV light for \num{10} sec (\textit{i.e.}, test for photo-bleaching) and examined with \num{490} nm the effect of that UV light exposure on the absorbance of this model drug. Next, scaffolds containing the drug molecules were transferred into syringes having exactly \num{500} µL of PBS and injected into \num{6}-well plates. Finally, the supernatants containing the released drug molecules were collected, and their absorbance was examined at \num{490} nm.

\subsection*{Acknowledgments}
This work was funded by a donation from Credit Suisse to the ETH Foundation to create a new chair for Robotics at ETH Zurich. The authors thank Susanne Freedrich from the ETH Phenomics Center for providing the blood samples and Karsten Kunze from the ETH Imaging Facility (ScopeM) for acquiring the electron microscopy images. O.Y. thanks Holcim Stiftung Wissen for the postdoctoral fellowship. \textbf{Author contributions:} O.Y. and R.K.K. designed the research. R.K.K. supervised the research. O.Y., F.M.T., M.F., and A.B. performed the experiments, analyzed the data, and generated the figures. O.Y., F.M.T., M.F., A.B., and R.K.K. wrote the manuscript.

\bibliography{scibib}
\bibliographystyle{Science}

\clearpage
\subsection*{FIGURES \& FIGURE CAPTIONS}

\begin{figure}[H] 
\includegraphics[width=16cm]{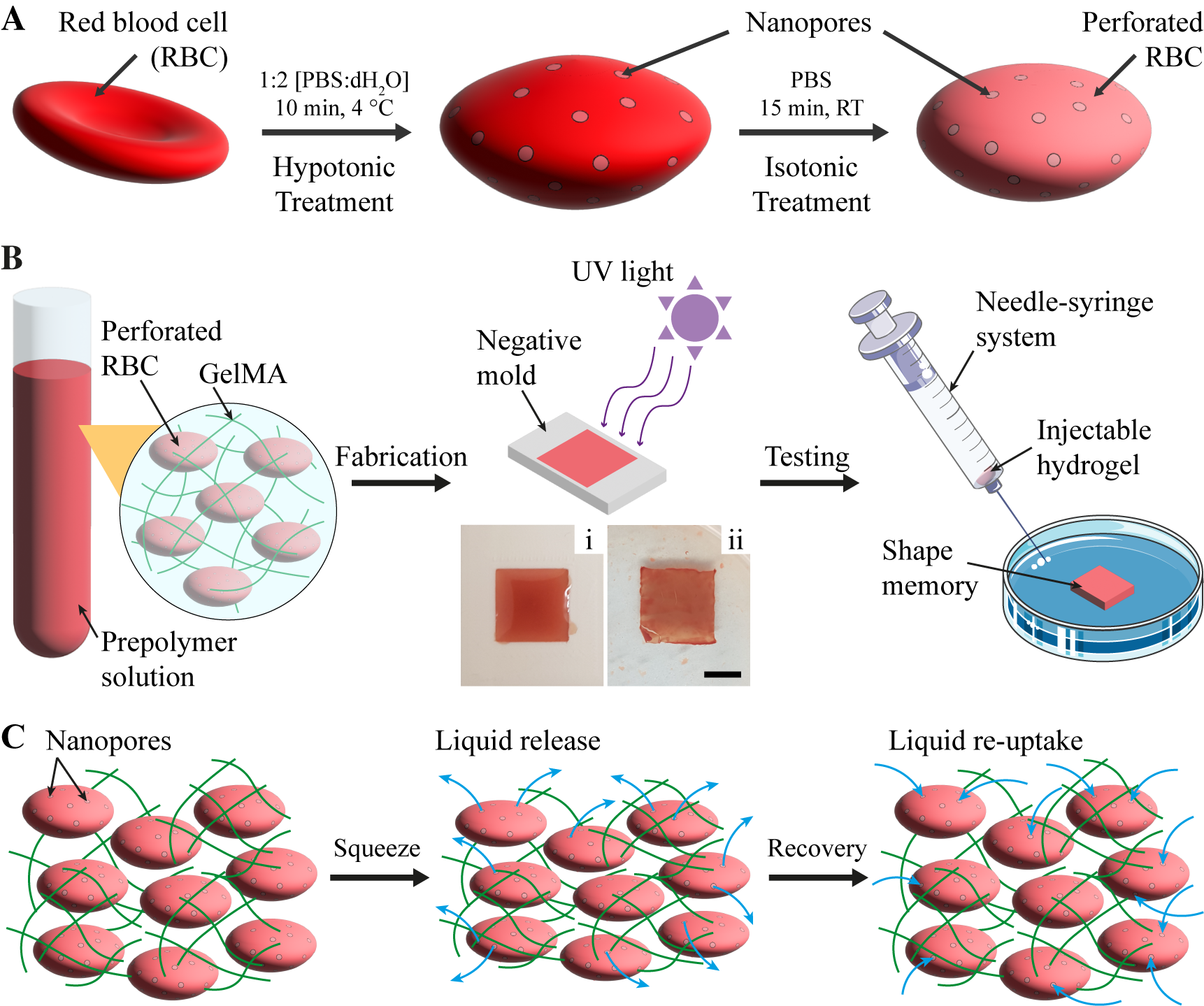}
\centering
\end{figure}
\noindent{\bf Fig. 1.} \textbf{The overview of the study.} \textbf{A)} The modified red blood cells (RBCs) with nanopores on their cell membranes were prepared using the hypotonic-isotonic treatment procedure. \textbf{B)} The prepolymer solution of injectable hydrogel scaffolds was formulated by homogeneously mixing the perforated RBCs with a photocrosslinkable polymer. The prepolymer-RBC solution was molded into 3D-printed structures \textbf{(i)} and crosslinked with a \num{405} nm ultraviolet (UV) light \textbf{(ii)}. Finally, the injectability and shape recovery of the fabricated scaffolds were tested by using conventional needle-syringe systems. \textbf{C)} The mechanism of injectability. The modified RBCs had nanopores on their cell membrane and acted as “space-holders” inside the hydrogel network. Under the compression, liquid stored inside the modified RBCs releases into the surrounding environment through the nanopores causing the hydrogel matrix to shrink. When the compression is removed, the modified RBCs re-uptake liquid from their surrounding environment and the hydrogel matrix returns its original shape. Blue arrows indicate the release and the re-uptake of the liquid through nanopores. 

\clearpage
\begin{figure}[H] 
\includegraphics[width=16cm]{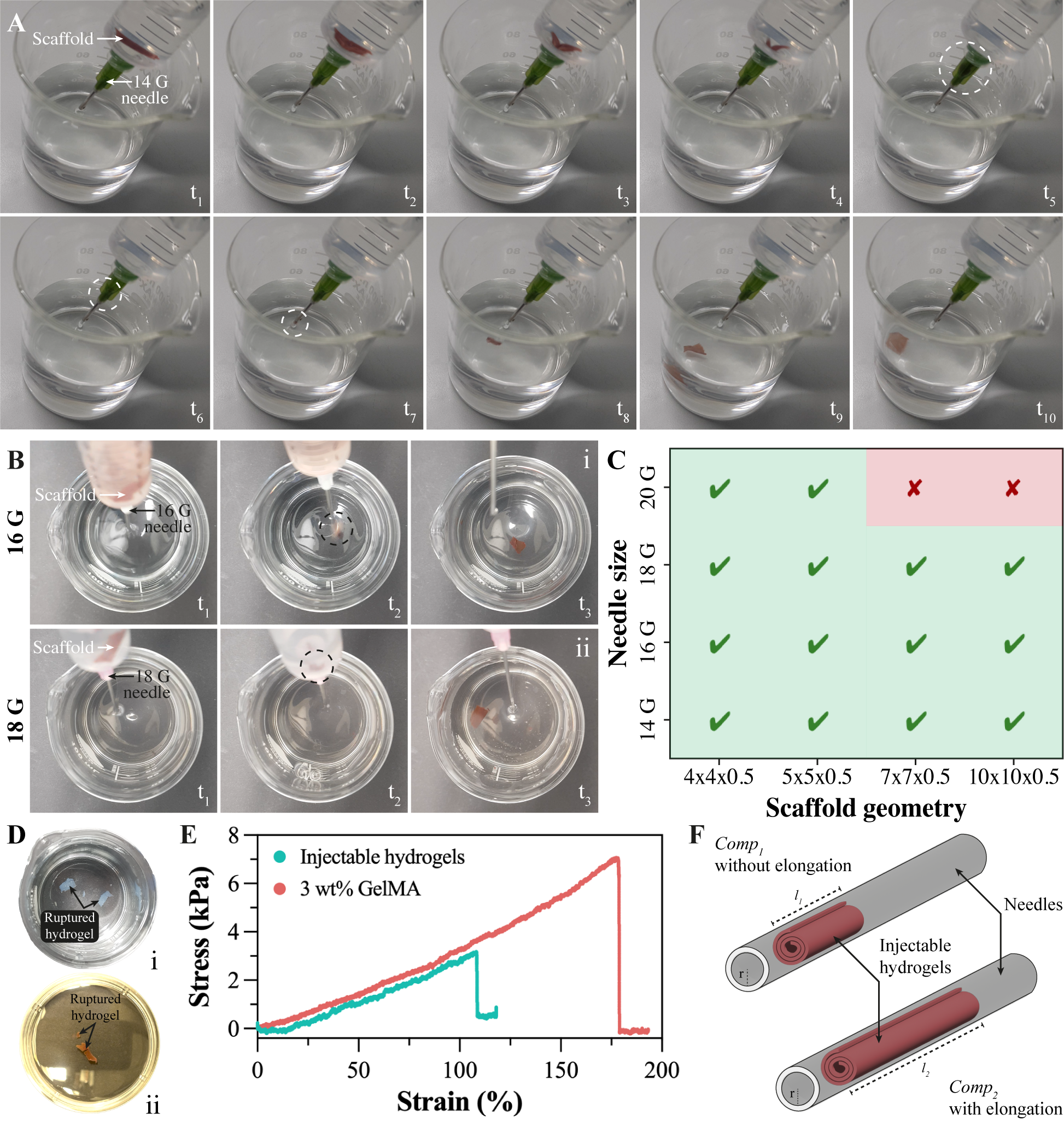}
\centering
\end{figure}
\noindent{\bf Fig. 2.} \textbf{The syringe injectability and mechanical characterization of hydrogels.} \textbf{A)} Injection of a \qtyproduct{10 x 10 x 0.5} {\mm} hydrogel scaffold with a \num{14} G needle; t$_1$ before injection, t$_2$-t$_7$ deformation while passing through the needle, and t$_8$-t$_1$$_0$ shape recovery after injection. White dashed circles show the hydrogel's position inside the needle. \textbf{B)} Injection of a \qtyproduct{10 x 10 x 0.5} {\mm} hydrogel scaffold with \textbf{(i)} \num{16} G and \textbf{(ii)} \num{18} G needles; t$_1$ before injection, t$_2$ deformation while passing through the needle, and t$_3$ shape recovery after injection. Black dashed circles show the hydrogel's position inside the needle. \textbf{C)} Injectability of hydrogel scaffolds having different geometries from \num{14}, \num{16}, \num{18}, and \num{20} needles. \textbf{D)} The rupture of \qtyproduct{10 x 10 x 0.5} {\mm} hydrogel scaffolds composed of either \textbf{(i)} only \num{3} wt\% GelMA or \textbf{(ii)} \num{3} wt\% GelMA with $\sim$\num{.5d9} unmodified RBCs/mL following their injection through \num{14} G needles. \textbf{E)} Strain-stress relation of the injectable hydrogels and control group I (\textit{i.e.}, hydrogels composed of only \num{3} wt\% GelMA). \textbf{F)} The assumed states of injectable hydrogel scaffolds (\textit{i.e.}, with and without elongation) inside needles that are used for the compression percentage calculations.

\clearpage
\begin{figure}[H] 
\includegraphics[width=16cm]{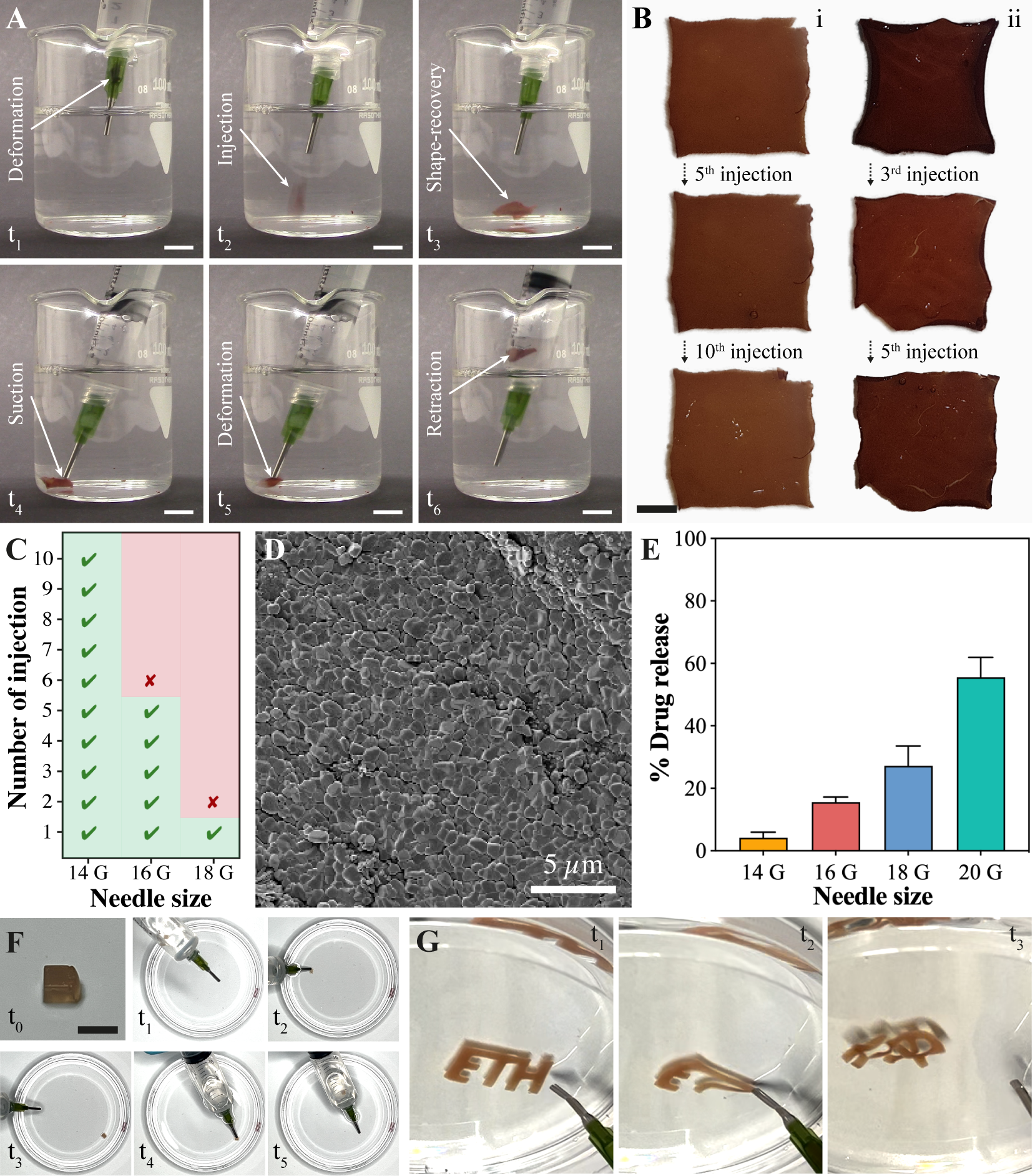}
\centering
\end{figure}
\noindent{\bf Fig. 3.} \textbf{The retraction and reusability of injectable hydrogels for concentrated therapeutic delivery.} \textbf{A)} The simultaneous injection and retraction of a \qtyproduct{10 x 10 x 0.5} {\mm} hydrogel scaffold with a \num{14} G needle. The scaffold deformed and squeezed through the needle (t$_1$) and was injected into a liquid environment (t$_2$). Then, it instantaneously recovered its original shape (t$_3$) by up-taking liquid from the surrounding environment. After that, it was aspirated with the needle-syringe system for retraction (t$_4$) and it started to deform at the tip of the needle (t$_5$). Finally, it passed into the syringe and again rapidly recovered its morphology (t$_6$). Scale bars represent \num{10} µm. \textbf{B)} The reusability of \qtyproduct{10 x 10 x 0.5} {\mm} hydrogel scaffolds following their injections through \textbf{(i)} \num{14} G and \textbf{(ii)} \num{16} G needles. Scale bar represents \num{2} µm. \textbf{C)} The number of injections of \qtyproduct{10 x 10 x 0.5} {\mm} hydrogel scaffolds through \num{14}, \num{16}, and \num{18} G needles. \textbf{D)} Scanning electron microscopy image of a injectable hydrogel scaffold following its three times injection through \num{14} G needle. \textbf{E)} Percentage of drug release from \qtyproduct{5 x 5 x 0.5} {\mm} hydrogel scaffolds following their single injection through \num{14}, \num{16}, \num{18}, and \num{20} G needles. \textbf{F)} The injection and retraction of \qtyproduct{3 x 3 x 3} {\mm} cubic hydrogel scaffold with a \num{14} G needle. The scaffold was properly crosslinked as a cube (t$_0$). Then, it was injected into a liquid environment with (t$_1$-t$_3$). Finally, it was retracted (t$_4$) and recovered inside the syringe (t$_5$). Scale bar represents \num{5} µm. \textbf{G)} The retraction (t$_1$-t$_2$) and re-injection (t$_3$) of a large hydrogel scaffold with a \num{14} G needle. Re-injected scaffold is scrambled but intact.

\clearpage
\subsection*{SUPPLEMENTARY INFORMATION}

\begin{figure} [H] 
\centering
\includegraphics[width=7.5cm]{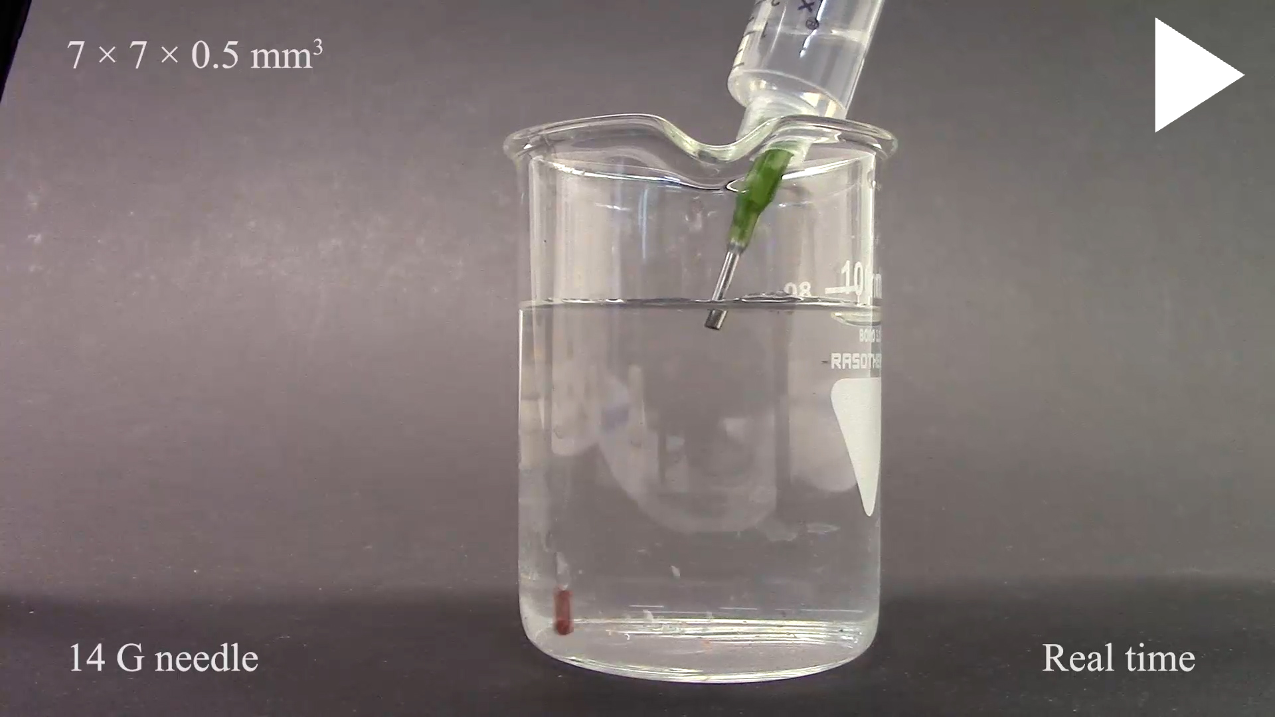}
\caption{\textbf{The syringe-injectability characterization of hydrogel scaffolds.} Four different geometries (\textit{i.e.}, \qtyproduct{10 x 10 x 0.5} {\mm}, \qtyproduct{7 x 7 x 0.5} {\mm}, \qtyproduct{5 x 5 x 0.5} {\mm}, and \qtyproduct{4 x 4 x 0.5} {\mm}) were examined by using \num{14}, \num{16}, \num{18}, and \num{20} G needles.}
\label{vid:video_s1}
\end{figure}

\begin{figure} [H] 
\includegraphics[width=7.5cm]{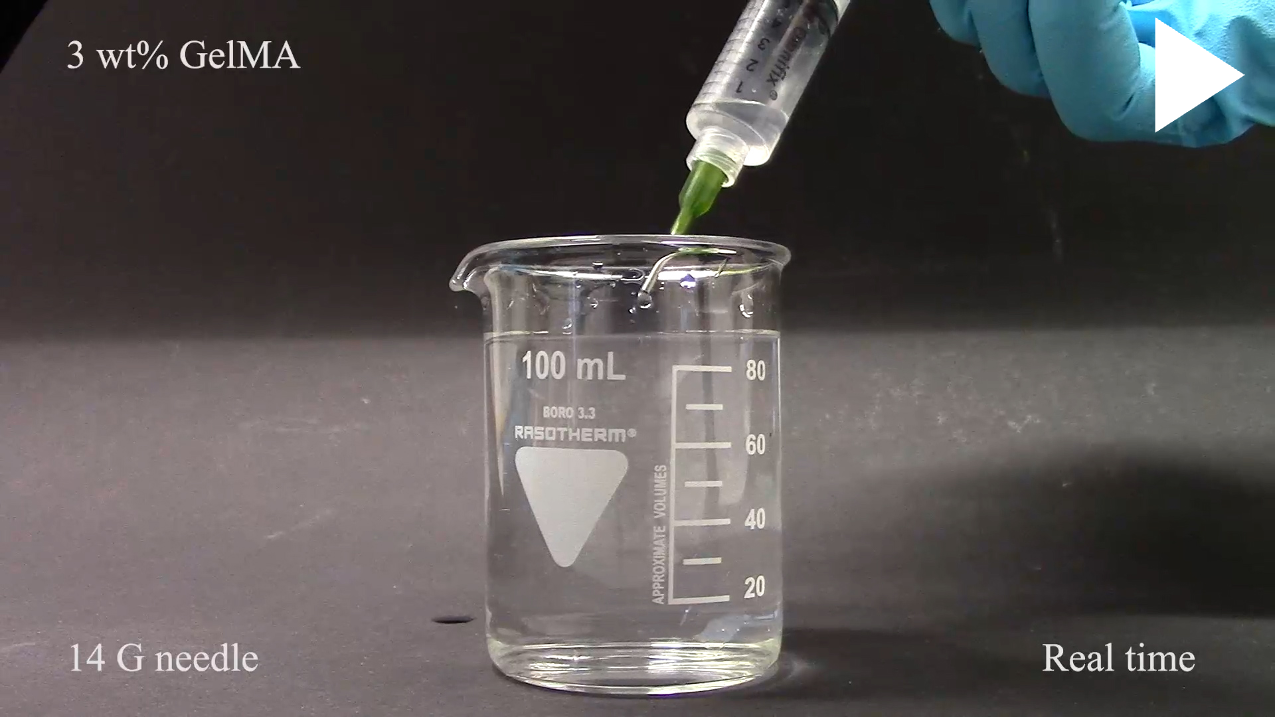}
\centering
\caption{\textbf{The injectability of hydrogel scaffold composed of only \num{3} wt\% GelMA.} The largest hydrogel scaffold (\textit{i.e.}, \qtyproduct{10 x 10 x 0.5} {\mm}) was examined with a \num{14} G needle (control group I).}
\label{vid:video_s2}
\end{figure}

\begin{figure} [H] 
\includegraphics[width=7.5cm]{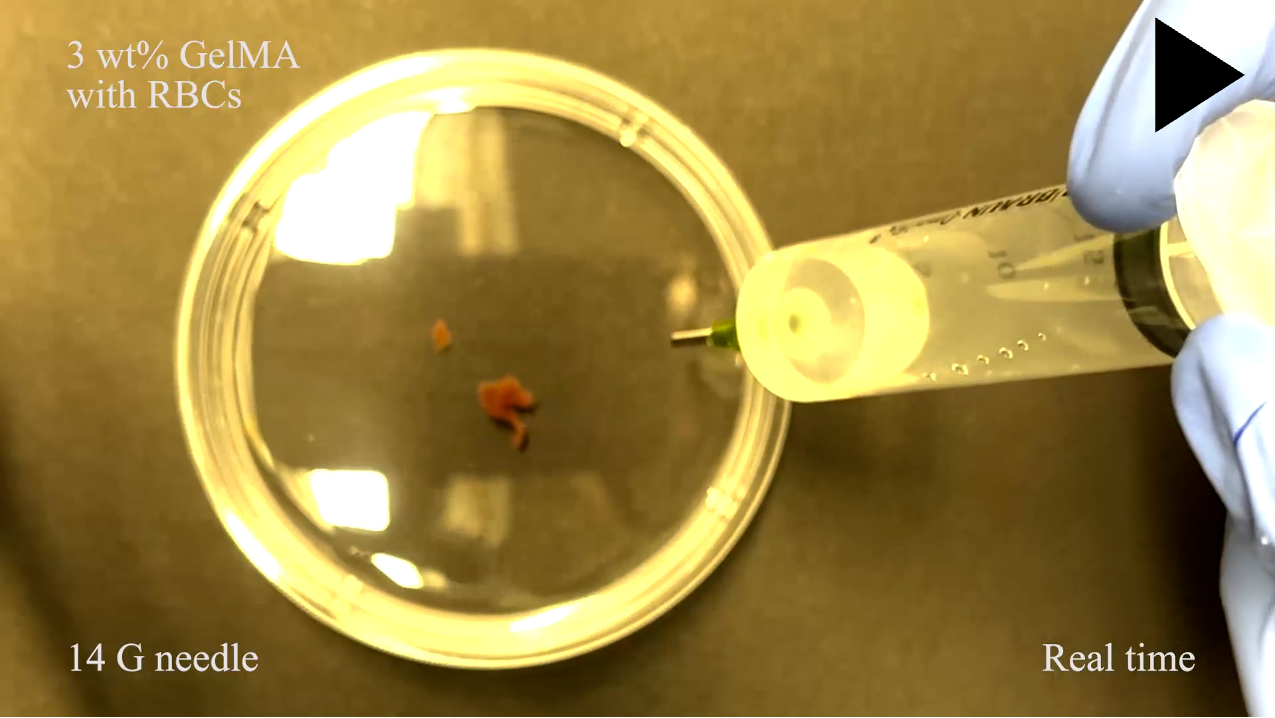}
\centering
\caption{\textbf{The injectability of hydrogel scaffold composed of \num{3} wt\% GelMA and unmodified RBCs.} The largest hydrogel scaffold (\textit{i.e.}, \qtyproduct{10 x 10 x 0.5} {\mm}) containing $\sim$\num{.5d9} unmodified RBCs/mL was examined with a \num{14} G needle (control group II).}
\label{vid:video_s3}
\end{figure}

\clearpage
\begin{figure} [H] 
\includegraphics[width=7.5cm]{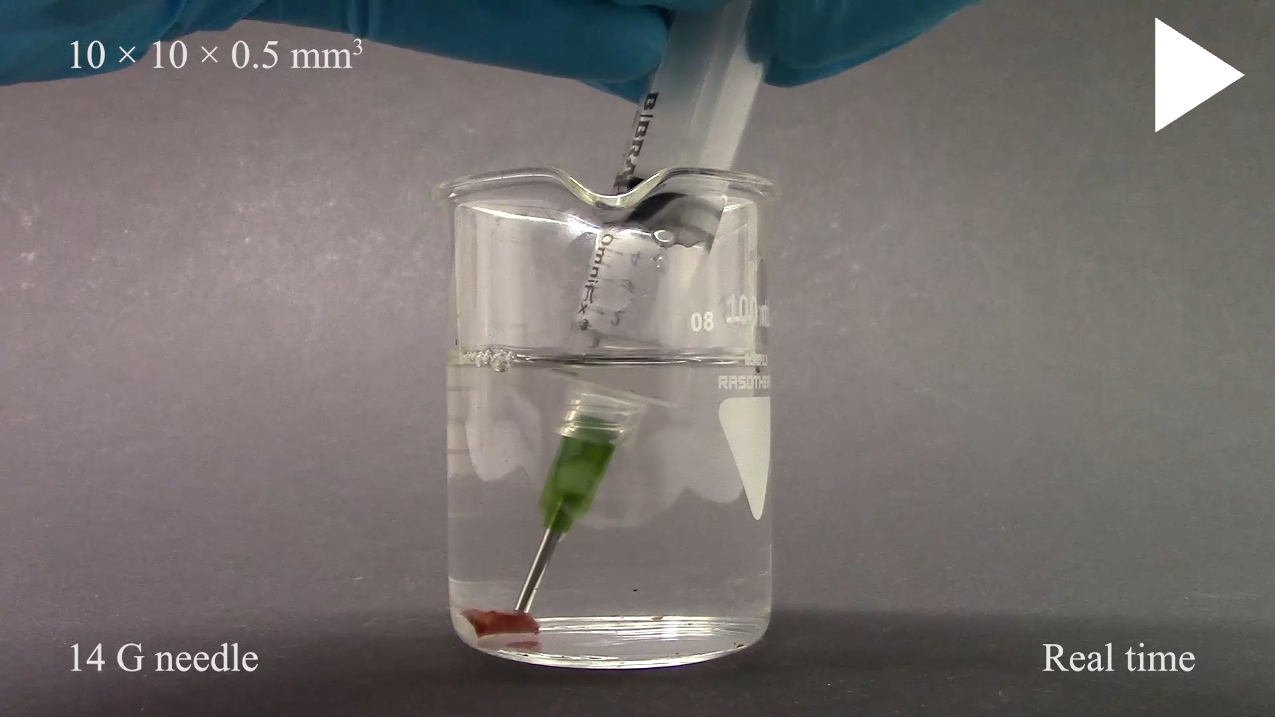}
\centering
\caption{\textbf{The retraction of hydrogel scaffolds with \num{14} and \num{16} G needles.} The retraction of the largest syringe-injectable hydrogel scaffolds (\textit{i.e.}, \qtyproduct{10 x 10 x 0.5} {\mm}) was examined by using \num{14} and \num{16} G needles.}
\label{vid:video_s4}
\end{figure}

\begin{figure} [H] 
\includegraphics[width=7.5cm]{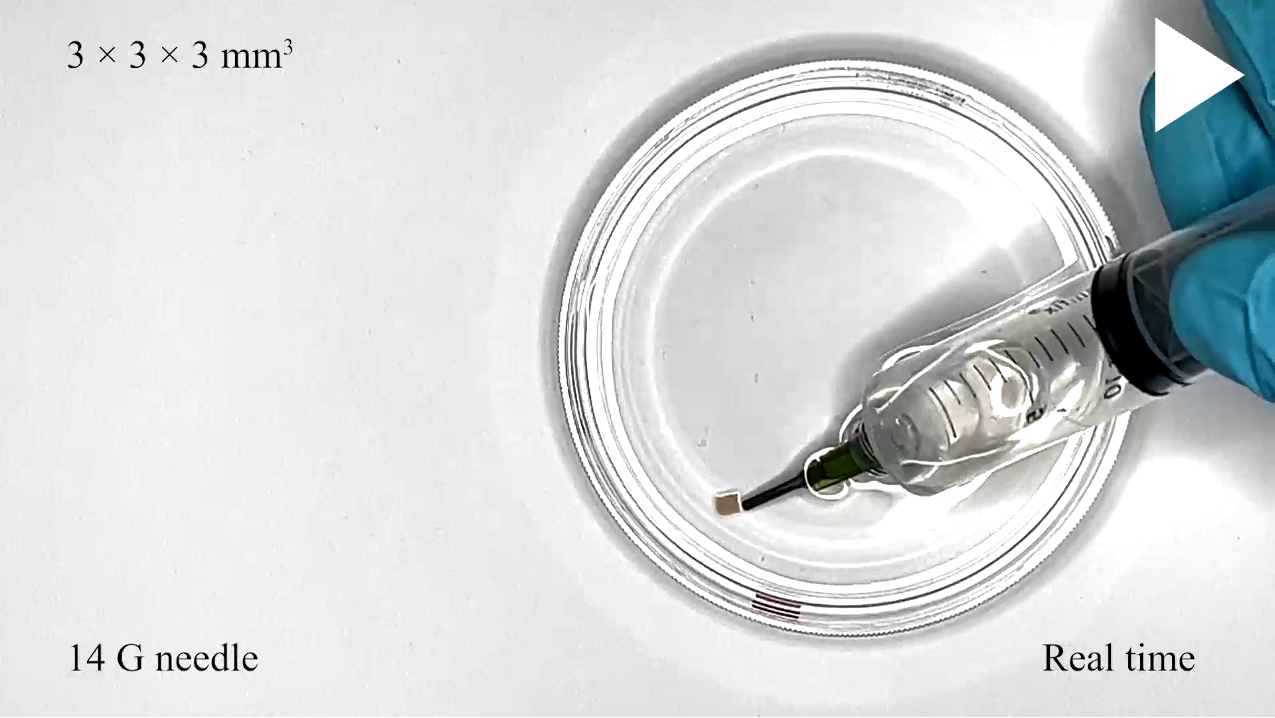}
\centering
\caption{\textbf{The injection and retraction of hydrogel scaffolds with different geometries.} A cubic and a large hydrogel scaffold with edges prone to damage were examined by using \num{14} needles.}
\label{vid:video_s5}
\end{figure}

\end{document}